\pdfoutput=1 
\documentclass[9pt,twocolumn,twoside]{opticajnl}
\journal{optica} 
\setboolean{shortarticle}{false}
\setboolean{displaycopyright}{false} 
\usepackage{lineno}
\usepackage{amsmath,amssymb}
\usepackage{siunitx} 
\usepackage{braket}
\usepackage{xcolor} 
\makeatletter
\let\oldGin@setfile\Gin@setfile
\renewcommand{\Gin@setfile}[3]{}
\makeatother

\title{Automated Vector-Scanning Spectroscopy for Large-Scale Characterization of Single Quantum Emitters}

\author[1,2]{William Eshbaugh}
\author[2,3]{Ashish Chanana}
\author[2,4]{Edgar Perez}
\author[2,3]{Junyeob Song}
\author[2]{Craig R. Copeland}
\author[1]{Sulaiman Al Ghadani}
\author[1]{Daniel McBride}
\author[1]{Prasiddha Siwakoti}
\author[5,2] {Maria Carolina Volpato}
\author[6]{Armando Rastelli}
\author[5]{Saimon Filipe Covre da Silva}
\author[1]{Ignacio Segovia-Dominguez}
\author[7]{Sadhvikas Addamane}
\author[2,4]{Kartik Srinivasan}
\author[1]{Edward B. Flagg}
\author[2,*]{Marcelo Davanco}

\affil[1]{West Virginia University, Morgantown, West Virginia, USA}
\affil[2]{National Institute of Standards and Technology (NIST), Gaithersburg, Maryland, USA}
\affil[3]{Theiss Research, La Jolla, California, USA}
\affil[4]{Joint Quantum Institute (JQI), University of Maryland, College Park, Maryland, USA}
\affil[5]{Universidade Estadual de Campinas (UNICAMP), Campinas, Brazil}
\affil[6]{Institute of Semiconductor and Solid State Physics, Johannes Kepler University, Linz, Austria}
\affil[7]{Center for Integrated Nanotechnologies (CINT), Sandia National Laboratories, Albuquerque, New Mexico, USA}
\affil[*]{Corresponding author: marcelo.davanco@nist.gov}
\dates{}

\begin{abstract} 
The inherent spatial randomness and broad spectral heterogeneity of epitaxial quantum dots (QDs)---one of the most mature classes of solid-state quantum emitters---remains a major obstacle to their scalable deployment in integrated photonic quantum technologies. Overcoming this challenge requires deterministic fabrication strategies capable of precisely aligning nanophotonic structures with high-quality emitters, which in turn demands efficient and automated single-QD characterization. Despite substantial progress in optical measurement techniques, a platform capable of autonomous, data-efficient, and sufficiently versatile characterization of single quantum dots at the chip scale remains lacking. Here, we introduce an automated cryogenic measurement platform that combines wide-field photoluminescence imaging with vector-stage-scanning confocal spectroscopy to enable high-throughput, chip-scale targeted optical characterization of individual QDs. Using this platform, we automatically acquire photoluminescence data from thousands of GaAs/AlGaAs QDs on a single chip. We demonstrate how this extensive dataset enables identification of high-performance emitters for future deterministic device fabrication, while simultaneously revealing statistical trends across the QD ensemble. By uniting data-efficient targeted measurements with scalable automation, our platform establishes a foundation for large-scale quantum photonic integration and the high-throughput characterization framework needed to accelerate materials optimization.
\end{abstract}

\begin{document}
\maketitle
\thispagestyle{empty} 

\makeatletter
\let\Gin@setfile\oldGin@setfile
\makeatother

\pagestyle{plain}

\begingroup
\renewcommand{\thefootnote}{}%
\stepcounter{footnote}
\footnotetext{This document has not been peer reviewed but has been cleared by NIST for release.}%
\endgroup

\section{Introduction}
Solid-state quantum emitters constitute key building blocks for emerging photonic quantum technologies~\cite{Aharonovich2016}. They can serve as bright, on-demand sources of flying qubits for photonic quantum simulation~\cite{wang_boson_2019,maring_versatile_2024} and linear optical quantum computing~\cite{ding_high-efficiency_2025}, as sources of entanglement for quantum communication~\cite{basso_basset_quantum_2021,vajner_quantum_2022} and quantum networking~\cite{anderson_gigahertz-clocked_2020,yu_telecom-band_2023}, as well as quantum information registers~\cite{appel_many-body_2025} and nonlinear optical elements operating at the single-photon level~\cite{javadi_single-photon_2015,tomm_photon_2023}. Among the various material systems explored, epitaxial III–V semiconductor quantum dots (QDs) represent the most technologically advanced class of solid-state quantum emitters, having benefited from more than three decades of extensive research and development~\cite{Michler2017}. Typically grown by molecular beam epitaxy (MBE) or metal-organic vapor phase epitaxy (MOVPE), they have enabled many of the milestone demonstrations that underpin quantum photonic technologies~\cite{wang_boson_2019,maring_versatile_2024,ding_high-efficiency_2025,basso_basset_quantum_2021,vajner_quantum_2022,anderson_gigahertz-clocked_2020,yu_telecom-band_2023,appel_many-body_2025,javadi_single-photon_2015,tomm_photon_2023}. The maturity of III–V epitaxy and device processing has furthermore facilitated the incorporation of single QDs into photonic micro- and nanostructures, enabling exploration of cavity quantum electrodynamics (QED)~\cite{lodahl_interfacing_2015} and integration within both monolithic~\cite{dietrich_gaas_2016,uppu_quantum-dot-based_2021} and heterogeneous or hybrid photonic integrated circuits~\cite{elshaari_-chip_2017,davanco_heterogeneous_2017,kim_hybrid_2017,katsumi_transfer-printed_2018,chanana_ultra-low_2022,larocque_tunable_2024}. Nevertheless, realizing scalable integration of reproducible, high-performance single-QD devices within programmable and multifunctional photonic circuits~\cite{wang_integrated_2020} remains an open challenge, the resolution of which could have a transformative impact on photonic quantum technologies. 

A primary challenge is the limited control over the location and spectral properties of individual QDs that is afforded by currently preferred self-assembled growth modes such as Stranski-Krastanov (S-K), droplet epitaxy (DE) and local droplet etching (LDE)~\cite{da_silva_gaas_2021,heindel_quantum_2023}. Achieving optimal device performance requires strict spatial and spectral alignment between single QDs and the confined optical modes of nanophotonic structures. This, in turn, demands deterministic targeting of individual QDs and lithographic patterning with total alignment uncertainties well below \qty{100}{\nano\meter} to attain reasonably high fabrication yields~\cite{copeland_traceable_2024}. Conventional fabrication methods, with no active targeting of emitters, are inherently low-yield due to the low probability of finding a device that contains not only a well-positioned, spectrally compatible QD, but also one of sufficient quality (e.g., narrow linewidth, high brightness, and minimal spectral diffusion). Significantly lower yields are expected if two or more devices with QDs having identical properties (e.g., the same emission energies for specific transitions) are required. As a result, deterministic fabrication approaches are preferred, where only the most favorable QDs are selected for integration. However, characterizing thousands to tens of thousands of emitters is not scalable manually, making an automated spectroscopy and localization platform essential. It is worth noting that, compared to other solid-state quantum emitters such as color centers in various hosts~\cite{sutula_large-scale_2023} and small organic molecules~\cite{toninelli_single_2021}, epitaxial quantum dots (QDs) are mesoscopic entities that exhibit a substantially broader distribution of optical properties within a single sample. Indeed, the inhomogeneous broadening of epitaxial QD ensembles typically spans tens of nanometers~\cite{grim_scalable_2019}, roughly 3 to 5 orders of magnitude larger than the lifetime-limited emission linewidths and at least an order of magnitude larger than that observed in color-center systems. This increased spectral variability places specific demands on large-scale optical characterization strategies. While localization-only approaches and recent site-templated QD growth techniques address the challenge of spatial alignment in targeted~\cite{pregnolato_deterministic_2020} and conventional array-based fabrication~\cite{gaur_buried-stressor_2025,zhang_-chip_2022}, respectively, modern scalable integration demands automated, high-throughput optical characterization platforms to efficiently identify and localize thousands of emitters.

Several experimental platforms have demonstrated automated spectroscopy and localization capabilities for large-scale quantum emitter characterization, primarily to improve the yield and throughput of deterministic nanofabrication. These methods can generally be categorized by their approach to spatial and spectral parallelization, each presenting unique tradeoffs between measurement capabilities and efficiency~\cite{gao_optical_2015}. Highly parallelized techniques offer the highest theoretical throughput but impose strict hardware or spectral limitations. For example, snapshot hyperspectral imaging (HSI)~\cite{liu_super-resolved_2024} enables fully parallelized spatial and spectral inference by reconstructing emission wavelengths from unique point spread function (PSF) fingerprints. However, this approach requires carefully engineered distributed-Bragg-reflector (DBR) cavities, limiting the range of photonic structures that can be subsequently fabricated around the QDs. Wide-field photoluminescence excitation (PLE) microscopy offers an alternative by combining parallel spatial imaging with serial spectral tuning~\cite{sutula_large-scale_2023}. While highly effective for color-center systems with intrinsically narrow spectral distributions, applying wide-field PLE to QD ensembles is technically challenging and inefficient due to intrinsically broad inhomogeneous spectral distribution. To accommodate the broad spectral distribution of QDs without relying on specialized sample geometries, alternative characterization platforms often involve exhaustive serial spatial-spectral mapping. These hyperspectral approaches typically record a photoluminescence (PL) or cathodoluminescence (CL) spectrum at every spatial coordinate---either by rastering the stage~\cite{sinclair_hyperspectral_2006} or beam~\cite{murali_phoqupy_2026, wijitpatima_bright_2024, barua_deterministic_2025}, or by scanning a magnified wide-field PL image across a spectrometer slit~\cite{buchinger_deterministic_2025-1}. Often optimized for basic spectral screening, these approaches are data inefficient for sparse emitter ensembles, where only a fraction of pixels contain useful emitter spectral information. This is sufficient for spectral alignment to photonic cavities, but for experiments involving deeper levels of single-emitter characterization---such as time-correlated single-photon counting or multidimensional parameter sweeps--- this overhead can make acquisition times and data storage prohibitive in these platforms. To circumvent this data inefficiency, alternative systems have been developed that utilize raster-scanning purely for raw intensity mapping using single-photon counters, and subsequently measuring potential emitters one-by-one using a beam vector-scanning technique for spectroscopy~\cite{liddlewesolowski_deterministic_2026}. Although data efficient, generating intensity maps via serial raster-scanning is temporally inefficient compared to parallel wide-field imaging, and to the best of our knowledge has not been completely automated for chip-scale measurements. Importantly, a targeted confocal measurement capability is required to perform post-fabrication device measurements---identifying cavity modes, quantifying radiative decay enhancement, or measuring single-photon correlations, for example.

In this work, we introduce an experimental platform that resolves these bottlenecks, enabling both high-throughput single-emitter spectroscopy for deterministic fabrication and detailed population-scale characterization in a framework that remains compatible with post-fabrication device testing. The setup combines wide-field PL imaging for parallel emitter localization with a vector-stage-scanning confocal spectroscopy system for targeted characterization of single QDs at cryogenic temperatures (<~\qty{4}{\kelvin}). By explicitly navigating to a subset of pre-selected emitters, the system avoids the substantial overhead associated with the exhaustive acquisition of spectral data from every pixel in a field of view. For context, we estimate a multidimensional HSI scan of a single field can produce nearly \qty{60}{\giga\byte} of data, whereas our targeted approach can acquire the exact same spectral information using less than \qty{150}{\mega\byte}---over a \qty{99}{\percent} reduction in raw data footprint (see Section 6 in the SI for throughput details). This selective acquisition strategy preserves the experimental accessibility and extensibility of traditional single-emitter spectroscopy while fully automating chip-scale measurements, enabling flexible operation across regimes: from population-level characterization of hundreds of emitters per field of view to selective, emitter-specific measurements focused on device-relevant candidates.

\begin{figure*}[h]
    \centerline{
    \includegraphics[width=2\columnwidth]{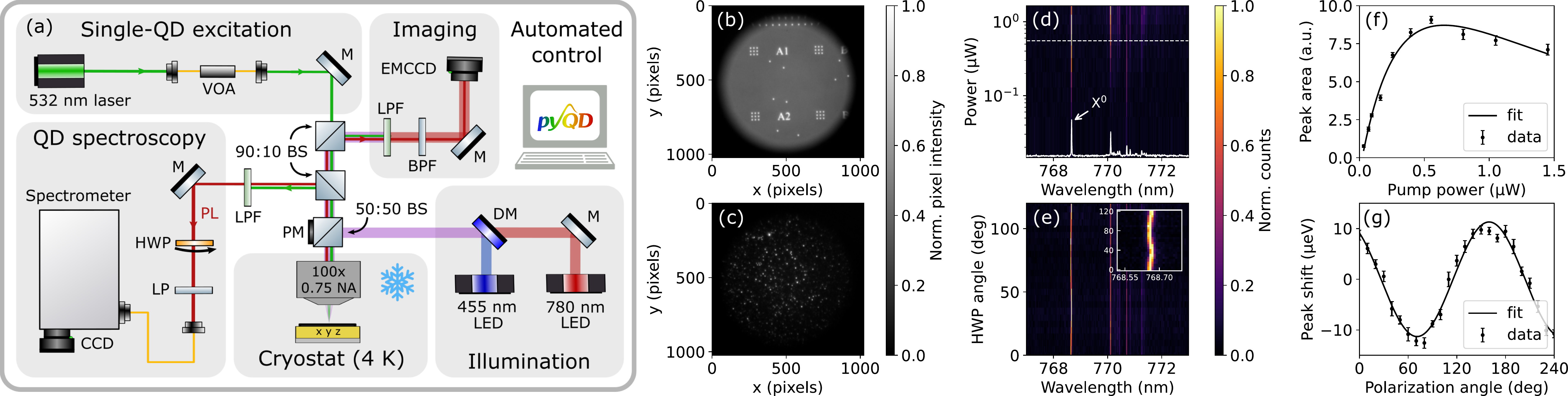}}
	\caption{
    (a) Experimental setup. BS: Beam splitter (reflection:transmission); M: mirror; DM: dichroic mirror. EMCCD: electron multiplying charge-coupled device; VOA: variable optical attenuator. HWP: half-wave plate; LP: linear polarizer; B/LPF: band/long-pass filter; PM: optical power meter. Exchangeable BPFs in front of the EMCCD allow imaging different spectral bands of the QD PL. Single QDs are addressed using a \qty{532}{\nano\meter} CW laser and VOA for power control. An automated rotating HWP enables polarization-resolved PL measurements. LPFs reject excitation light from collection and imaging paths. (b,c) Wide-field fiducial and PL images, respectively. The metallic dots within the field serve as \textit{pseudo} emitters intended for future localization validation experiments. (d) Power-dependent spectra of a single QD. The spectrum overlay (white) corresponds to the slice taken near maximum neutral exciton ($\mathrm{X}^0$, indicated) PL intensity at a pump power of approximately \qty{0.5}{\micro\watt} (dashed line). (e) Polarization-dependent spectra of the same QD. Inset: ROI around the $\mathrm{X}^0$ line. (f) Power series analysis of the $\mathrm{X}^0$ peak chosen from (d). (g) $\mathrm{X}^0$ fine-structure splitting (FSS) analysis of the data in (e). Least-squares fit models for the data in (f) and (g) are described in the SI. Vertical bars are Type A (statistical) standard uncertainties from the fit standard errors (normal approximation, 46 degrees of freedom, $\approx$\qty{68}{\percent} confidence intervals). 
	} 
	\label{fig:setup_diagram}
\end{figure*}
This enables straightforward access to power-dependent saturation behavior, polarization-resolved fine-structure splitting, voltage-dependent Stark shifts, time-resolved lifetime measurements, or confocal PLE scans---measurements that can become at least cumbersome if not impossible in alternative approaches. Although this stage-scanning approach does not maximize raw throughput in the same manner as previously described alternatives, it provides a configurable trade-off between throughput and characterization depth, offering reduced data volume, improved experimental flexibility and extensibility, and direct access to detailed single-emitter physics. We anticipate that our selective measurement framework can also be leveraged for automated testing of fully fabricated single QD devices, and that the ideal platform for targeted, serial single-emitter characterization and parallel localization will combine wide-field imaging capabilities with vector-scanning confocal spectroscopy.

We employ our setup to automatically characterize an ensemble of LDE GaAs QDs with emission wavelengths between \qty{760}{\nano\meter} and \qty{800}{\nano\meter} and observe a significant variation in spectral quality of individual QDs across an overall population of thousands. Such variation, which cannot be deduced from imaging data alone, can significantly impact the fabrication yield of nanophotonic devices around targeted QDs, beyond just spectral misalignment. By imaging and inspecting statistical trends within different spectral bands, we observe an apparent wavelength-dependent bimodal distribution of QD spectral quality through a linewidth analysis, which is consistent with clusters identified in the behavior of power-dependent spectra (see Section 7 in the SI). We further demonstrate that promising QDs can be preferentially selected while avoiding emitters that are unsuitable for fabrication. This information can be leveraged to improve the yield of high-quality emitters identified for downstream deterministic fabrication. We anticipate the wealth of intimate single-emitter data will enable even more sophisticated QD characterization and pre-selection approaches through clustering or machine learning~\cite{corcione_machine_2024}, which can advance understanding of QD population(s), highlighting potential beyond deterministic fabrication; e.g., optimizing QD growth processes. In the following sections, we describe our automated spectroscopy setup in detail and provide insights to its utilization toward deterministic single-QD device integration and statistical characterization of QD populations.

\section{Automated Spectroscopy Platform}\label{sec:platform}
\subsection{Hardware description}
Fig.~\ref{fig:setup_diagram}(a) shows a schematic of our experimental setup. Epitaxial QD samples are measured in a \qty{4}{\kelvin} closed-cycle cryostat featuring a top lid window that provides optical access to the cold space. An infinity-corrected, apochromatic 0.75 NA microscope objective mounted inside the vacuum chamber allows stable, high-magnification imaging of the sample with a field-of-view of approximately \qty{60}{\micro\meter}. The samples are mounted on a three-axis, resistively encoded nanopositioner stack with a maximum lateral range of \qty{5}{\milli\meter}. A two-color wide-field illumination scheme is employed to image single QD PL alongside fiducial markers, an established technique for localization microscopy~\cite{sapienza_nanoscale_2015,liu_nanoscale_2021}. In this technique, fiducial markers provide a coordinate reference for localizing single QD PL images, which appear as point-spread functions (PSFs) within an imaging field. The fiducial markers consist of 3 by 3 square arrays of metallic circles with a diameter of \qty{1}{\micro\meter} and \qty{2.5}{\micro\meter}, and are located at the four corners of a \qty{50}{\micro\meter} square imaging field as shown in Fig.~\ref{fig:setup_diagram}(a). This design enables PSF registration using only 2D Gaussian fitting, avoiding the need for more complex edge-detection algorithms used in alternative fiducial patterns~\cite{liu_nanoscale_2021, buchinger_deterministic_2025-1}. Various grids of such imaging fields were produced on a single \qty{10}{\milli\meter} by \qty{10}{\milli\meter} chip.

A \qty{455}{\nano\meter} LED is used to provide incoherent, above-bandgap excitation of the QDs in their III-V semiconductor host, while a \qty{780}{\nano\meter} LED is used for reflection-based imaging of metallic fiducial markers on the sample surface. The latter wavelength is chosen to overlap with the expected QD emission, to minimize potential chromatic aberration effects. LED powers can be manually adjusted for optimal image quality.
\begin{figure*}[t]
    \centerline{
    \includegraphics[width=1.95\columnwidth]{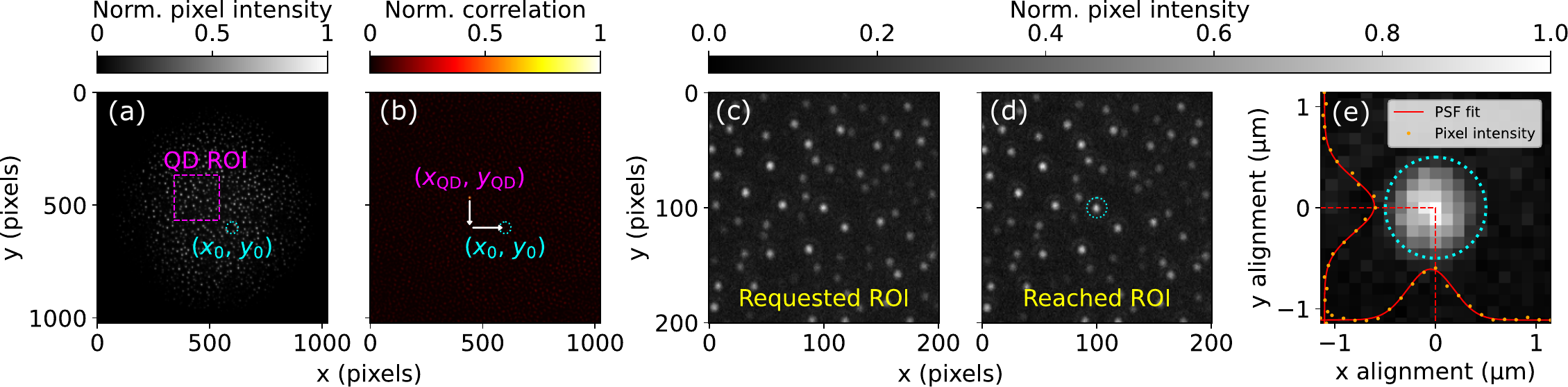}}
	\caption{ 
	The cross-correlation-based navigation process. (a) Wide-field QD PL image of a representative field of view. ($x_0$, $y_0$) is the position of the fixed confocal spot, and the magenta box illustrates a typical template extracted around a PSF. (b) Normalized, positive-only cross-correlation between the DoG-filtered template and image shown in (a). The correlation maximum of 0.99 corresponds to the target PSF position ($x_{\mathrm{QD}}$, $y_{\mathrm{QD}}$), which is used to compute pixel displacement vector components (white arrows) to the confocal spot. The pixel displacement is scaled to a discrete number of steps and the target PSF is iteratively guided to its destination. (c, d) The initial \textit{requested} ROI from (a), and the final \textit{reached} ROI extracted around the confocal spot after navigation, respectively, revealing successful navigation of the selected emitter PSF to the confocal point. (e) The target PSF excited by an above-band laser. Slices of the fitted PSF (through the confocal origin, indicated by dashed lines) are shown along each axis, from which a total alignment offset of \qty{52.9}{\nano\meter} is estimated, with a combined standard uncertainty of \qty{3.6}{\nano\meter} between the target PSF centroid and the confocal center. The uncertainty is Type A, propagated from spatial-fit standard errors (354 degrees of freedom, normal approximation, $\approx$68~\% coverage). Pixel alignment offsets are converted to micrometers using a scale factor estimated from system calibration~\cite{copeland_traceable_2024}. The cyan circle represents an estimated \qty{1}{\micro\meter} laser spot size (not to scale in (a-d)).
	}
	\label{fig:cross_correlation}
\end{figure*}
A back-illuminated silicon electron-multiplied charged-coupled device (EMCCD), with a resolution of $1024\times1024$ pixels (\qty{13}{\micro\meter} pixel size), is used to image both the QD PL and the fiducial markers. While simultaneous imaging of both is possible, alternating between the two LEDs allows us to obtain separate images of the QDs and fiducials with independently optimized acquisition parameters for good signal-to-noise ratio (SNR). A selectable band-pass filter placed in front of the camera enables observation of different spectral bands within the full QD population, reducing PSF density in the image and improving detection, selection, and navigation. A \qty{532}{\nano\meter} continuous wave (CW) laser is used to incoherently excite single emitters at a fixed confocal excitation and PL collection spot. PL from single QDs is collected through the internal objective, sent through a linear polarization analyzer, coupled into a single-mode fiber, and analyzed with a grating spectrometer equipped with a liquid-N$_2$-cooled silicon CCD. The polarization analyzer, consisting of an automated rotating half-wave plate (HWP) and linear polarizer (LP), enables polarization-resolved PL measurements. An electronic variable optical attenuator (VOA) enables the collection of power-dependent spectra. Automated instrument control code is available~\cite{midas_doi}.

\subsection{Template-matching-based stage navigation}\label{sec:automation}
In order to automate selective QD spectroscopy, each emitter must be reliably positioned at the fixed confocal excitation and collection spot. To accomplish this, we use a template-matching method based on normalized image cross-correlations~\cite{lewis_fast_2001,walt_scikit-image_2014}) to locate the same PSF in updated PL images obtained after stage motion, iteratively displacing the target PSF until it reaches the pixel coordinate of the fixed confocal spot (see Fig.~\ref{fig:cross_correlation}). This navigation approach is purely image-based: pixel displacement vectors are computed from a correlation map to guide stage motion, eliminating the need for encoded stage positions, which can be an unreliable navigation method at the scales required to accurately and unambiguously position emitters at the confocal spot (see Section 5 in the SI). The navigation process is detailed in the next two paragraphs.

PSFs are first detected in the wide-field PL image and subsequently filtered analytically by proximity, symmetry, and brightness following the approach detailed in Section 2 of the SI, which reduces the number of candidate emitters to visit. These pre-selection filters can be relaxed to select all available emitters within a field, or restricted to limit selection to a handful of isolated, bright, or dim PSFs per field. A nearest neighbor traversal order is assigned to visit the next closest PSF, which avoids any inefficient navigation paths that could arise from default ordering of emitters by PSF pixel intensity. Fig.~\ref{fig:cross_correlation} illustrates the subsequent template matching process that enables navigation between emitters. In this process, a region-of-interest (ROI) image is first extracted around a selected PSF to serve as a template to be matched against the full wide-field PL image (Fig.~\ref{fig:cross_correlation}(a)). The maxima in the resulting correlation map corresponds to instances of the template image present in the wide-field image (Fig.~\ref{fig:cross_correlation}(b)). The template ROI is centered on the target PSF and includes some of the surrounding PL structure to provide a unique PSF constellation for correlation. The ROI size plays an important role in navigation fidelity and depends on PSF density, intensity, and noise. If the ROI is too small, for example, and contains only the target PSF, many similar local maxima will appear in the correlation image. Careful tuning of ROI size and imaging exposure time (i.e., signal-to-noise ratio) is required to preserve template matching accuracy while maintaining reasonable navigation efficiency. Additionally, to help prevent false-positives arising from intensity-dominated correlations, gamma intensity compression and difference-of-Gaussians (DoG) filtering are applied to both the template and wide-field PL image, which effectively prioritizes structural correlation over raw pixel intensity correlation (see Section 5 in the SI for details). The correlation maximum occurs where the template best overlaps with the wide-field image, yielding the target PSF coordinate via a weighted centroid calculation over a local ROI. A displacement vector (in pixels) is then calculated between the detected emitter position $(x_{\mathrm{QD}}, y_{\mathrm{QD}})$ and the target confocal spot $(x_0, y_0)$. To convert this image-space displacement into physical stage movement, we employ an adaptive navigation scheme. First, the magnitude of the displacement vector dictates the piezo positioner's drive voltage according to a manually defined distance schedule (i.e., larger pixel displacements use higher voltages). The chosen voltage then determines a scale factor (inversely proportional to the driving voltage) to convert the raw pixel displacement into a discrete number of piezo steps along each axis. By dynamically coupling both the step amplitude (voltage) and the step count to the target distance, the stage takes large initial movements that progressively become finer as the PSF approaches the confocal spot. This schedule is empirically tuned to ensure alignment converges within a few iterations while effectively suppressing oscillatory overshoot at sub-pixel precision. Fig.~\ref{fig:cross_correlation}(e) shows the target PSF under laser excitation, demonstrating sub-\qty{100}{\nano\meter} alignment accuracy. Once positioned at the target PSF, a final template is extracted around the confocal spot to verify the positioned ROI matches the requested one using a verification cross-correlation (see Figs.~\ref{fig:cross_correlation}(c-d)). The maximum correlation value is assigned as a confidence score, which can be used during experiments or in post-processing to flag potential false positives (see Section 5 in the SI). Single-QD measurements, such as the representative data shown in Fig.~\ref{fig:setup_diagram}(d-e), are then performed. Importantly, we have not observed any significant stage drift, and the target PSFs remained well within the estimated confocal spot area during the measurements performed in this work. All data analysis (such as shown in Fig.~\ref{fig:setup_diagram}(f-g)) is performed separately to maintain automation efficiency and role separation. However, relevant analyses could be incorporated at runtime to tune the excitation wavelength for a PLE scan and perform lifetime or photon-purity measurements, for example.

\begin{figure*}[t!]
    \centerline{
    \includegraphics[width=2\columnwidth]{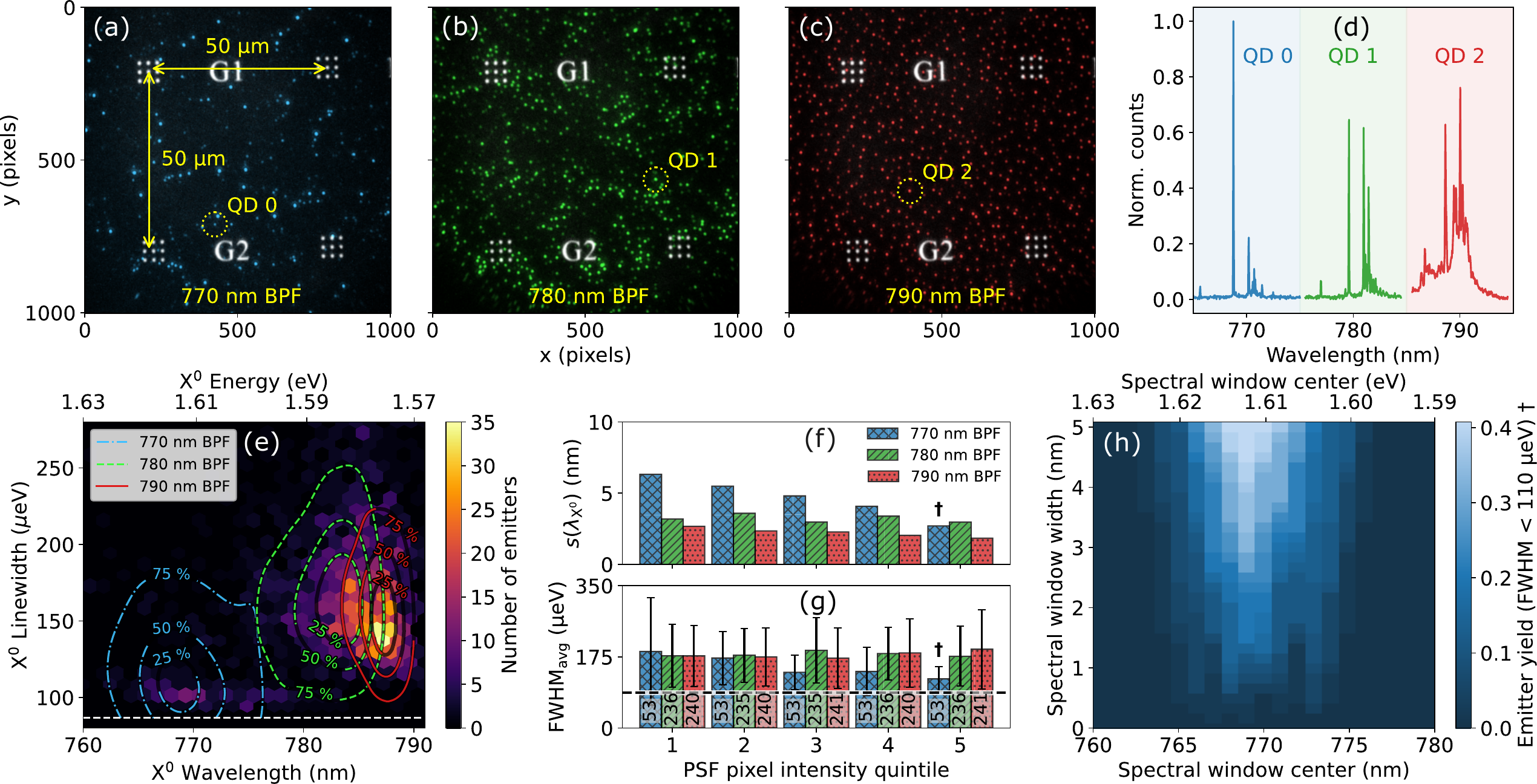}}
	\caption{
    	QD population statistics. (a-c) False color QD PL images with fiducial overlay, after filtering with \qty{770}{\nano\meter} (blue), \qty{780}{\nano\meter} (green), and \qty{790}{\nano\meter} (red) band-pass filters (BPFs), respectively. Yellow circles denote selected QDs with spectra shown in (d). (d) Spectra of QDs annotated in (a-c). (e) Density plot of neutral exciton ($\mathrm{X}^0$) linewidth and emission wavelength, revealing an apparent bimodal distribution. Contours indicate highest-density regions containing the indicated percentages of emitters measured with each BPF, and the dashed white line indicates the \qty{40}{\pico\meter} (\qty{82}{\micro\electronvolt}) spectrometer resolution. (f) Sample standard deviation $s(\lambda_{\mathrm{X}^0})$ of the $\mathrm{X}^0$ peak wavelength within PSF pixel intensity quintiles to illustrate the spectral spread within each bin. (g) Average FWHM within each PSF pixel intensity quintile. Here, quintiles correspond to \qty{20}{\percent} increments of increasing PSF pixel intensity. The values annotated on each bar in (g) indicate the number of emitters in each bin, and the dashed line indicates the spectrometer resolution. Vertical bars in (g) indicate the standard deviation of each bin's FWHM distribution to illustrate spectral uniformity within each bin ($\dagger$ indicates the bin chosen for yield estimate in (h)). (h) Fraction of ``good'' QDs that fall within spectral windows of varying width and center wavelength. QDs are selected from the top \qty{20}{\percent} of PSF intensities in the \qty{770}{\nano\meter} dataset ($\dagger$) and exhibit $\mathrm{X}^0$ linewidth FWHM~<~\qty{110}{\micro\electronvolt}.
        }
	\label{fig:populations}
\end{figure*}

To automate chip-scale single-QD measurements, the stage must also reliably navigate to unique \qty{50}{\micro\meter}$\times$\qty{50}{\micro\meter} QD imaging fields comprising a larger grid across the sample. Field-to-field navigation is performed using a coarse translation to the target position based on readout of encoder coordinates. For non-encoded positioners, we expect this process can still be used with careful stage calibration, where a discrete number of steps can be taken to  position the stage on a new field of view with sufficient accuracy. In either case, on samples patterned with fiducials for emitter localization experiments, as shown in Fig.~\ref{fig:populations}(a)-(c), a field-centering step is performed by template-matching after coarse motion, where a template containing a single set of the $3\times3$ metallic fiducials is used to locate the four field corners and guide their centroid toward the center of the imaging field. We emphasize that this fine-centering step can be bypassed if fiducials are not present, which enables navigation and characterization of single emitters across an arbitrary number of user-defined grids. To compensate for macroscopic sample tilt across millimeter-scale regions, an automated through-focus sweep is performed at each field of view. Using either the QD PL or fiducial images, the stage height is adjusted to maximize a gradient-based image sharpness metric~\cite{krotkov_focusing_1988, sun_autofocusing_2004}. At this optimal focal plane, separate images of the QD PL and fiducial markers are acquired for localization and pre-selection. Furthermore, we anticipate this template-matching navigation framework can be adapted for in-cryostat testing of fully fabricated devices, offering some potential for improved measurement throughput and repeatability.

To characterize the throughput of our system, we focus primarily on navigation time, which directly reflects how fast individual emitters and different fields can be visited. For total experimental runtime, spectral measurements will be the primary bottleneck, which scales linearly with the number of emitters and will heavily depend on experimental parameters (e.g., integration time, acquisitions per series, and series type). From the most recent navigation statistics, our system takes approximately \qty{10}{\second} to navigate between emitters and \qty{20}{\second} to navigate between fields of view, including the auto-focus step. A more detailed throughput discussion is provided in Section 6 of the SI.

\section{Quantum Dot Population Statistics}
We employed the system described above to characterize ensembles of individual GaAs QDs grown by the LDE epitaxy method across seven \qty{50}{\micro\meter} by \qty{50}{\micro\meter} fields. The investigated sample consisted of a \qty{10}{\milli\meter} by \qty{10}{\milli\meter} GaAs chip with the following epitaxial layers: \qty{500}{\nano\meter} of Al$_{0.75}$Ga$_{0.25}$As; \qty{5}{\nano\meter} GaAs; \qty{140}{\nano\meter} Al$_{0.4}$Ga$_{0.6}$As; and \qty{5}{\nano\meter} GaAs. The GaAs QDs were grown in the center of the Al$_{0.4}$Ga$_{0.6}$As film, \qty{70}{\nano\meter} away from the top and bottom GaAs cap layers. The top GaAs cap is present to prevent oxidation of the Al$_{0.4}$Ga$_{0.6}$As device layer. The Al$_{0.75}$Ga$_{0.25}$As film was grown to serve as a sacrificial layer for future fabrication of suspended nanophotonic devices. In this case, the bottom GaAs cap layer also protects the active structure from oxidation. Fiducial markers as described above, shown in Fig.~\ref{fig:populations}(a-c), were produced on top of the sample by electron-beam lithography and metal lift-off. We automatically collected wide-field QD PL and fiducial field images along with power-dependent spectra for all detected PSFs in three different imaging spectral windows.

Automatic characterization was performed primarily to determine the spectral quality of individual QDs, informing pre-selection for chip-scale mapping intended for targeted fabrication of photonic nanostructures, where identical emitters are preferred for repeatable device fabrication. Generally, LDE QDs exhibit a consistent spectral structure, featuring a well-defined neutral exciton ($\mathrm{X}^0$) peak alongside a cluster of lower-energy peaks associated with charged and excited excitonic species~\cite{huber_single-particle-picture_2019} (as depicted in Fig.~\ref{fig:setup_diagram}(d)). Accordingly, the full width at half maximum (FWHM) of the $\mathrm{X}^0$ peak was used as a simple proxy for emitter quality, with narrower linewidths indicating higher suitability for device integration. As shown later, the large statistical dataset revealed broader trends in the QD population. 

We initially imaged the QD ensemble PL through a \qty{700}{\nano\meter} long-pass filter, which revealed an average QD density of approximately \qty{0.15}{\per\micro\meter\squared} with a sample standard deviation of \qty{0.01}{\per\micro\meter\squared}, which indicates the physical variation of QD density across seven fields of view. To reduce the imaged PSF density and minimize the potential for QD misidentification, we opted to image the QD population sequentially through \qty{10}{\nano\meter} band-pass filters (BPFs) centered at \qty{770}{\nano\meter}, \qty{780}{\nano\meter}, and \qty{790}{\nano\meter}. Fig.~\ref{fig:populations}(a)-(c) show typical PL images obtained from a single field, where spectrally-dependent QD density and clustering are evident. Representative spectra from emitters in each spectral band are shown in Fig.~\ref{fig:populations}(d). QDs found in the shortest-wavelength window are relatively sparse with isolated clusters, while clustering becomes less obvious in the longer-wavelength window, with a transition in the intermediate window. 

The distribution of $\mathrm{X}^0$ linewidth and emission wavelength, shown in Fig.~\ref{fig:populations}(e), reveals a large spread across the full ensemble. The ensemble appears to follow a bimodal distribution, with narrow linewidth QDs below approximately \qty{780}{\nano\meter} (\qty{1.59}{\electronvolt}), and broader linewidth QDs at longer wavelengths. Emitters within the \qty{770}{\nano\meter} BPF band, despite spanning a wide range of emission wavelengths and having a lower spatial density, predominantly exhibit near-resolution-limited linewidths (the \qty{40}{\pico\meter} (\qty{82}{\micro\electronvolt}) spectrometer resolution is indicated by the horizontal dashed lines in Fig.~\ref{fig:populations}(e,g)). In contrast, the longer-wavelength spectral bands contain a greater number of emitters, but a higher proportion display significantly broader linewidths. We hypothesize that this bimodality in spectral quality might arise from QDs grown on a template exhibiting a bimodal distribution of shallow and deep etched nanoholes, as reported by Heyn \textit{et al.} in Refs.~\cite{heyn_optical_2009, heyn_single-dot_2010} but requires further investigation to verify. We apply an unsupervised machine-learning algorithm and observe a similar wavelength-dependent bimodality in excitation-power-dependent spectral behavior (see Section 7 in the SI).

Motivated by the emitter pre-selection criteria mentioned in Section~\ref{sec:platform}\ref{sec:automation}, we analyzed spectral data with respect to PSF pixel intensity (the raw intensity taken from the PSF center), dividing the population into five percentile bins to identify population trends and improve the selection of narrow-linewidth emitters with similar emission wavelengths. Fig.~\ref{fig:populations}(f,g) shows the spread of emission wavelength and average linewidth across PSF intensity quintiles for each spectral band. The values provided here serve purely as descriptive metrics of the distributions within each bin. For the \qty{770}{\nano\meter} spectral band, the average $\mathrm{X}^0$ wavelength in the lowest PSF intensity bin is \qty{771}{\nano\meter} with a standard deviation of \qty{6.3}{\nano\meter} (this standard deviation represents the physical spread of emission wavelength, not the uncertainty of the mean). Intuitively, this large spread can be explained by the dim PSF selection, which effectively samples the tails of a broader QD spectral distribution selected by the \qty{770}{\nano\meter} BPF. Conversely, the average emission wavelength from the brightest PSF intensity bin is \qty{768}{\nano\meter} with a standard deviation of \qty{2.7}{\nano\meter}, showing a significant reduction in the inhomogeneous spectral distribution of the brightest QDs in that band. Similarly, we observe a reduction in the average peak FWHM from \qty{90.4}{\pico\meter} (\qty{188.5}{\micro\electronvolt}) to \qty{57.5}{\pico\meter} (\qty{120.9}{\micro\electronvolt}) along with a reduction in standard deviation of the FWHM distributions from \qty{62.8}{\pico\meter} (\qty{130.6}{\micro\electronvolt}) to \qty{14.2}{\pico\meter} (\qty{30.4}{\micro\electronvolt}), suggesting more uniform spectral behavior within this bin. Thus, pre-selecting emitters from the top \qty{20}{\percent} of pixel intensities can minimize both spectral spread and linewidth within the sub-population isolated by the \qty{770}{\nano\meter} BPF---desirable targets for nanophotonic integration. As evident in the \qty{780}{\nano\meter} and \qty{790}{\nano\meter} datasets, where there is a minimal change in wavelength spread and peak width relative to PSF intensity, emitter pre-selection criteria may not be defined as easily and alternative pre-selection methodologies should be explored. For instance, our system's automated capability to acquire spectra for thousands of individual emitters could facilitate the training and application of more advanced pre-selection or classification machine-learning models~\cite{corcione_machine_2024}.


We also performed a yield analysis to quantify how many spectrally proximate narrow-linewidth emitters are available within specified spectral intervals across the \qty{770}{\nano\meter} spectral band. We qualitatively label emitters featuring $\mathrm{X}^0$ linewidth FWHM~<~\qty{110}{\micro\electronvolt} (\qty{54}{\pico\meter}) as ``good'', which corresponds to the local minimum of a bimodal Gaussian fit to the linewidth distribution Fig. \ref{fig:populations}(e). Sweeping a variable-width sliding spectral window across the $\mathrm{X}^0$ emission wavelength distribution to count only the good emitters, we obtained the fractional yield plot in Fig. \ref{fig:populations}(h).
This analysis can provide valuable quantitative feedback to optimize deterministic device integration yield. For instance, device design can be tuned to target spectral regions with the highest density of high-quality emitters within a given spectral window, where the spectral window may correspond to, e.g., a narrow optical cavity resonance, or the (often limited) QD spectral tuning ranges afforded by the quantum-confined Stark effect or strain-based techniques \cite{sun_strain_2013, chen_wavelength-tunable_2024, grim_scalable_2019}.

When planning an automated chip-scale measurement intended for deterministic single QD device integration, we can estimate the number of emitters to measure in order to reliably find a target number of desirable QDs, $\tilde{k}$. Assuming independent visitations with a constant success probability $p$ based on our empirically observed fractional yield of narrow linewidth QDs, this process follows a binomial model (serving as a conservative upper-bound approximation for finite-population sampling without replacement). To determine the experimental sampling burden, $N$, required to guarantee finding at least $\tilde{k}$ suitable emitters with \qty{95}{\percent} certainty, we evaluate the corresponding cumulative distribution function (CDF). Considering that the observed fractional yield is limited by sampling statistics, a \qty{95}{\percent} binomial proportion confidence interval (BPCI) is calculated to provide upper and lower bound estimates for $p$~\cite{nist_sematech_ehandbook}. The probability limits are then applied to the CDF to establish best- and worst-case bounds for the required measurement burden $N$. Suppose we require $\tilde{k}=100$ ``good'' emitters within a \qty{2}{\nano\meter} spectral window centered on \qty{769}{\nano\meter}. Searching among the entire QD population for the \qty{770}{\nano\meter} BPF range results in an observed fractional yield of \qty{1.2}{\percent}. Using the Wilson score interval for the \qty{95}{\percent} BPCI~\cite{nist_sematech_ehandbook} to account for the yield uncertainty, our CDF calculation estimates that between 6540 and 14662 visitations are required to hit the $\tilde{k}=100$ target with \qty{95}{\percent} confidence. However, limiting the search to the top \qty{20}{\percent} brightest PSFs in the \qty{770}{\nano\meter} band (Fig.~\ref{fig:populations}(h)), with a corresponding fractional yield of approximately \qty{21}{\percent}, assumed to be uniformly distributed across all fields, the required measurement burden drops significantly. For this targeted approach, visiting between 340 and 966 QDs ensures the same \qty{95}{\percent} probability of success. This roughly 17-fold increase in search efficiency translates directly into faster measurements, reduced data overhead, and improved throughput.

\section{Conclusion and Outlook}
We have developed an automated vector-scanning confocal spectroscopy and wide-field imaging system capable of performing chip-scale spectroscopy and localization of single quantum emitters with the primary goal of enabling deterministic integration into on-chip photonic nanostructures. Beyond fabrication, such capability enables the acquisition of rich single-emitter luminescence spectrum datasets, which can provide valuable insight into QD populations to optimize growth processes and potentially train more sophisticated classification models to further improve pre- or post-selection of high-quality emitters. Our template-matching–based navigation enables completely encoder-free positioning of emitters within a given field of view, reducing operational complexity while maintaining sub-\qty{100}{\nano\meter} positioning accuracy to a fixed confocal spot. While our setup uses encoded position values for coarse inter-field movement, we expect this can be implemented for non-encoded systems through careful stage calibration and that template matching-based navigation could be extended to automate the positioning of fabricated devices and improve post-fabrication characterization throughput. We demonstrate that initial population-level measurements can inform emitter selection, for example by identifying optimal spectral bands and intensity thresholds for pre-selection. This enables efficient targeting of the highest-quality QDs for deterministic device integration. When combined with simple localization fiducials, the system further ensures spatial repeatability during fabrication, which is essential for achieving high-performance photonic devices where spectral alignment, brightness, and positional accuracy are critical.

While alternative automated spectroscopy schemes can provide more efficient parallelized population-scale screening, as outlined in the introduction, our selective characterization approach provides a balance between throughput and single-emitter measurement capabilities. Although spectroscopy is not parallelized in the same manner as QD localization, it supports both population-level screening and flexible characterization within the same framework and already implements several single-emitter optical characterization measurements such as fine-structure splitting and power-dependent saturation. Importantly, the underlying framework is readily extensible for more in-depth single-QD characterization methods such as PLE, dark-field resonance fluorescence~\cite{kuhlmann_dark-field_2013}, Stark-effect QD tuning, and time-resolved measurements. 

Overall, our system enables automated luminescence characterization of large numbers of single QDs in emitter ensembles, providing significant flexibility to allow measurements that balance throughput, precision, and richness of experimental data. We anticipate such capabilities will provide a practical path towards faster QD growth optimization cycles informed by detailed data sets, as well as for scalable, deterministic, high-yield fabrication of nanophotonic devices based on individual quantum emitters.

\noindent\textbf{Funding.}
K.S. acknowledges NSF Award 2531569 (National Quantum Virtual Laboratory: Quantum Computing Applications of Photonics). W. Eshbaugh acknowledges financial assistance under Award No. 70NANB23H257 from the U.S. Department of Commerce, National Institute of Standards and Technology. S. Al Ghadani acknowledges support from the National Science Foundation under award number 2327223. E. B. Flagg and D. McBride acknowledge support from the Office of Naval Research (ONR) under Award No. N00014-23-1-2611. A. Rastelli acknowledges support of the Austrian Science Fund (FWF) [10.55776/COE1, 10.55776/PIN4389523, 10.55776/FG5]. This work was performed, in part, at the Center for Integrated Nanotechnologies, an Office of Science User Facility operated for the U.S. Department of Energy (DOE) Office of Science. Sandia National Laboratories is a multimission laboratory managed and operated by National Technology \& Engineering Solutions of Sandia, LLC, a wholly owned subsidiary of Honeywell International, Inc., for the U.S. DOE’s National Nuclear Security Administration under contract DE-NA-0003525. The views expressed in the article do not necessarily represent the views of the U.S. DOE or the United States Government. 


\noindent\textbf{Disclosures.}
Generative AI (Google Gemini Pro 3.1) was used to assist in reviewing and editing the draft of this manuscript. The authors have reviewed, edited, and take full responsibility for the final content of the publication. This manuscript identifies certain commercial products to adequately describe an experimental process. Such identification does not imply recommendation or endorsement by the National Institute of Standards and Technology, nor does it imply that the products are the best available for the purpose.

\noindent\textbf{Data Availability Statement}
The instrument control code, data, and data analysis code underlying the results presented in this paper will be made available in the NIST Public Data Repository upon publication. DOIs will be provided at the proof stage.

\bibliography{references_Ned_v2,references_MD}

\end{document}


\maketitle

\section{Chip-scale Characterization}
\begin{figure}[h!]
    \centerline{
    \includegraphics[width=0.98\textwidth]{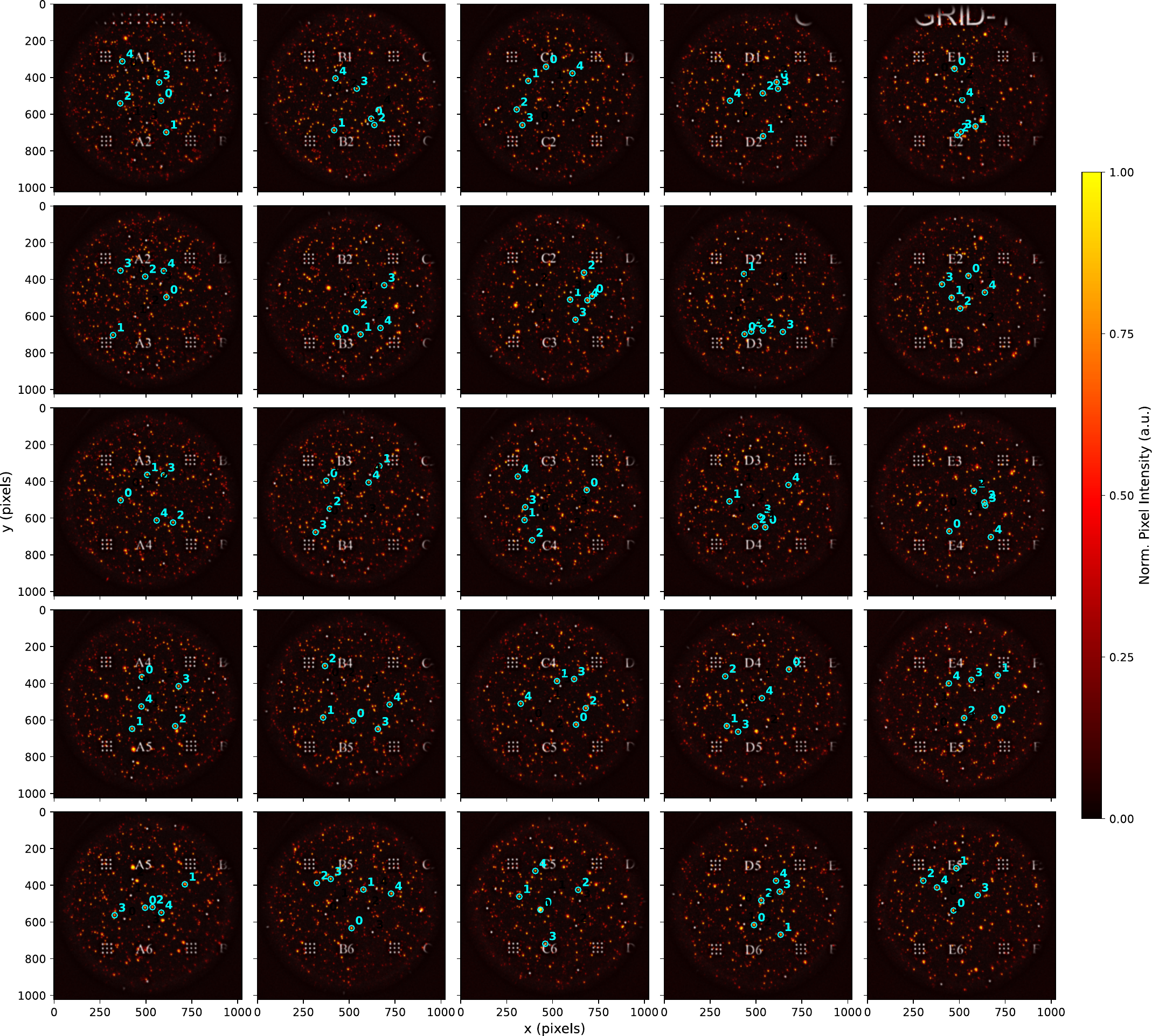}}
	\caption{
    Representative grid of merged fiducial and PL field images (\qty{50}{\micro\meter} between the centers of the corner fiducial marks) demonstrating field navigation consistency during chip-scale measurements. The color bar represents normalized pixel intensity of the PL, while the fiducials are shown in white. A typical sample includes several $9\times9$ grids, but a reduced $5\times5$ set is shown here for visibility. QDs (annotated in cyan) were selected following the process illustrated in Fig.~\ref{fig:SI_selection}, and their positions were registered to the fiducial marker coordinate frame for later targeted fabrication.
	} 
	\label{fig:SI_fields}
\end{figure}

To demonstrate chip-scale measurement capabilities, we performed automated imaging and spectroscopy of four sets of $9\times9$ grids of \qty{50}{\micro\meter}~$\times$~\qty{50}{\micro\meter} imaging fields on a single \qty{10}{\milli\meter}~$\times$~\qty{10}{\milli\meter} chip. A subset of such imaging fields is shown in Fig.~\ref{fig:SI_fields} for visualization. When targeting single QDs for deterministic device fabrication, selection of only a handful of emitters within an imaging field, as illustrated in Fig.~\ref{fig:SI_selection}, is in principle desirable for efficiency.

\section{Emitter Detection and Selection}
\begin{figure}[h]
    \centerline{
    \includegraphics[width=1\textwidth]{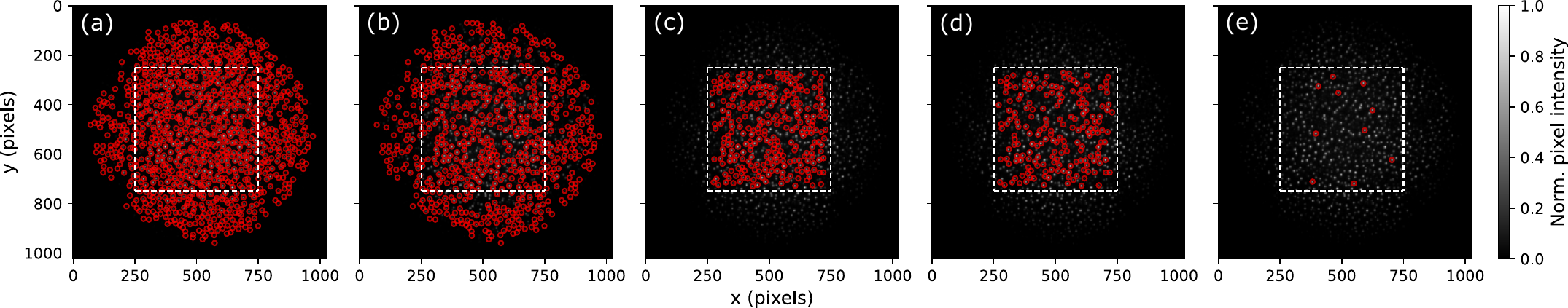}}
	\caption{
    PSF pre-selection filters. (a) Initial DoG blob detection of all possible PSF candidates. (b) PSFs remaining after application of a proximity filter with a separation threshold of 20 pixels. (c) Excluding emitters outside a defined field boundary, which is used to avoid cross-field measurements in the fiducial-defined grids. Remaining emitters are subsequently fit with a 2D Gaussian model that allows for asymmetry. (d) Symmetry filter selecting PSFs that satisfy a maximum ellipticity constraint (1.3 in this example, where ellipticity is defined as the ratio between the maximum and minimum 2D Gaussian width parameters). (e) A final selection of 10 emitters from a desired pixel intensity percentile, the top \qty{50}{\percent} brightest in this case.
	} 
	\label{fig:SI_selection}
\end{figure}

An example of the emitter detection and selection process is shown in Fig.~\ref{fig:SI_selection}. Point spread function (PSF) spots are first detected in the wide-field PL image using a standard Difference-of-Gaussians (DoG) blob detection method~\cite{walt_scikit-image_2014}). To reduce the number of candidate emitters to consider, an initial set of coarse filters select PSFs that fall within a defined field boundary and are spatially isolated by a defined minimum pixel separation from neighboring PSFs. Each PSF in the remaining set is fit with a 2D Gaussian model using a nominally single-PSF range-of-interest (ROI) to ensure robust and sub-pixel localization precision. A subsequent symmetry filter selects PSFs that satisfy a maximum ellipticity constraint (defined as the ratio between maximum and minimum 2D Gaussian width parameters), which helps remove any spatially unresolvable clusters of nearby emitters that were not caught by the proximity filter, and a final filter allows a specified number of dots to be requested from a pixel intensity range. These filters can be relaxed to select all available emitters within a field, or restricted to limit selection to a handful of isolated, bright or dim emitters per field. Once a set of emitters is selected, a nearest neighbor greedy traversal algorithm is used to assign traversal order. In other words, navigation will automatically visit the emitter that is closest to the destination coordinate rather than following the inefficient route produced by the default emitter ordering after selection (e.g., brightest to dimmest).

\section{Data Analysis}\label{sec:SI_data_analysis}
A parallelized analysis pipeline~\cite{midas_doi} was developed to efficiently process the large amount of spectral and localization data collected during automated measurements. For first-level spectral power series analysis, peaks are detected in individual slices of the series using a common peak detection algorithm~\cite{2020SciPy-NMeth}. These detected peaks are fed into a DBSCAN (density-based spatial clustering for applications with noise) algorithm~\cite{ester_density-based_1996,pedregosa_scikit-learn_2011}, to identify recurring peaks across the series. This measure avoids erroneous cosmic ray detection and enables peak fitting in a fixed set of wavelength windows across the series. Identified peaks are fit with standard Gaussian, Lorentzian, or Voigt profiles with basic linear or constant background terms, which are fed into a secondary analysis layer to extract derived quantities such as saturation power and fine-structure-splitting (FSS). For completeness, the models fitted to the data in Fig. 1 of the main text are described below, but the analysis pipeline is extensible and any user-defined model can be implemented.

The saturation curves are fit with a model modified from~\cite{hummel_efficient_2019} to include an empirical quenching term:
\begin{equation}
    I(P) = I_0(1-e^{-\frac{P}{P_{sat}}})e^{-\frac{P}{rP_{sat}}},
\end{equation}

where $I(P)$ is the integrated PL intensity for each detected peak at each excitation power $P$, $I_0$ is an overall intensity scale, $P_{sat}$ defines the characteristic power scale of saturation, and $r$ is a dimensionless parameter which sets the relative scale of the exponential quenching process.

To estimate FSS, the tracked $\mathrm{X}^0$ emission wavelength is fit with a simple sinusoid as a function of the HWP angle~\cite{seidl_statistics_2008, plumhof_experimental_2010, ollivier_three-dimensional_2022}:

\begin{equation}
    \lambda(\theta) = \lambda_0 + \frac{\Delta\lambda}{2} \sin(2\theta + \phi),
\end{equation}

where $\theta$ is the polarization angle, $\lambda_0$ is the mean emission wavelength, $\Delta\lambda$ is the peak-to-peak wavelength separation (i.e., fine-structure splitting), and $\phi$ is a phase offset that encodes dipole orientation relative to the reference frame defined by the HWP and linear polarizer alignment.

To account for the possibility that detected peaks originate from different quantum dots (QDs) within a single spectrum and mitigate misidentification of the neutral exciton ($\mathrm{X}^0$) peak (assumed to be the highest-energy isolated emission line from a single QD~\cite{huber_single-particle-picture_2019}), we select $\mathrm{X}^0$ from all detected peaks based on spectral proximity to the corresponding band-pass filter (BPF) window. If the highest-energy detected peak lies within the BPF window, it is selected as $\mathrm{X}^0$. Otherwise, we search for the highest-energy peak within a \qty{10}{\nano\meter} window centered on the nearest in-band detected peak, treating this cluster as belonging to the same QD. If no peaks fall within the BPF window, we instead select the detected peak closest in wavelength to the BPF center.

For statistical analysis, a single representative FWHM is obtained by computing an inverse-variance weighted mean of the $\mathrm{X}^0$ linewidths across the full spectral power series, where each linewidth is weighted by $1/\sigma_{\mathrm{FWHM}}^2$, with $\sigma_{\mathrm{FWHM}}$ the standard fit uncertainty. To avoid double counting emitters that appear in multiple BPF datasets due to spectral overlap between adjacent passbands, duplicate PSFs are identified across datasets using a \qty{200}{\nano\meter} proximity threshold and retained only in the dataset with the highest PSF intensity. This threshold corresponds to approximately $0.6\times$ the fitted PSF standard deviation and is chosen to be restrictive enough to avoid flagging spatially separate emitters as duplicates, ensuring that matched detections are consistent with originating from the same physical emitter. Prior to this comparison, a similarity transform is applied to align fiducial markers across images, ensuring consistent registration between spectral bands. A direct spectral comparison is complicated by variations in excitation and collection alignment arising from independent navigation during measurements within each BPF. For the PSF pixel intensity correlations in Fig.~\ref{fig:SI_intensity_corr}, no de-duplication was performed so that the intensity percentiles are computed from the full population and reflect independent measurements for each spectral band.

\section{Encoder-based Navigation Ambiguity}
\begin{figure}[h]
    \centerline{
    \includegraphics[width=1.0\textwidth]{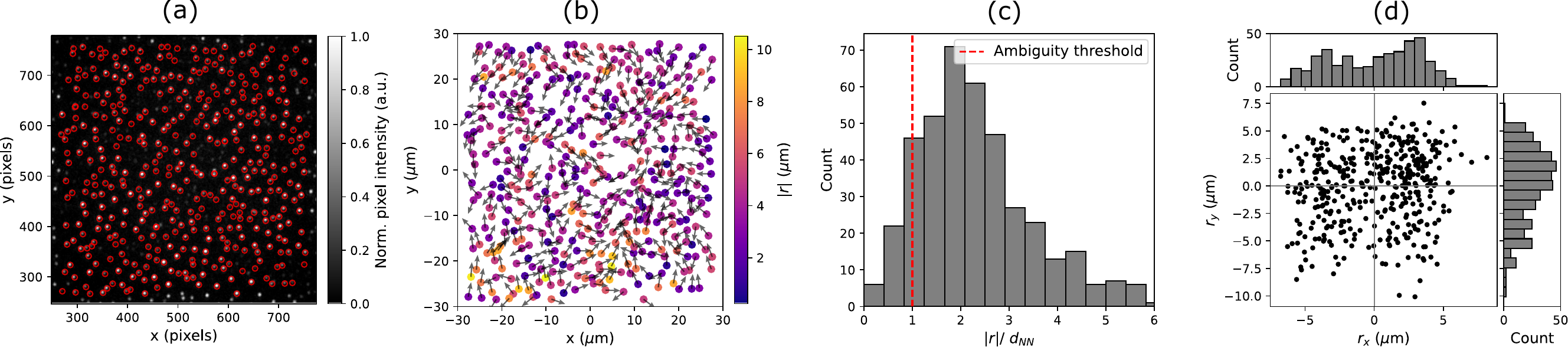}}
	\caption{
    (a) Cropped wide-field PL image showing the visited emitters (red circles). (b) Corresponding real-space map constructed from scaled image-based coordinates. Residual vectors are computed point-by-point between encoder and image-based positions and are shown as color-coded magnitudes with unit vectors indicating direction. The spatial arrangement of emitters is preserved under the coordinate transformation. (c) Distribution of residual magnitudes $|r|$ normalized by the mean nearest-neighbor spacing $d_{NN}$. The distribution indicates that residuals are comparable to or exceed the characteristic emitter spacing (red dashed line), leading to potential ambiguity in emitter identification. (d) Distribution of residual components $r_x$ and $r_y$, showing no significant directional bias.
    } 
	\label{fig:SI_encoder}
\end{figure}

To further illustrate the utility of our template-matching-based navigation approach, we compare single QD coordinates obtained from the resistive encoder readout of the piezoelectric positioning stages after visiting each QD with coordinates obtained from PSFs localized in the wide-field photoluminescence (PL) image (Fig.~\ref{fig:SI_encoder}(a)).
Image-based coordinates are first converted to micrometers via a scale factor obtained from a least-squares similarity transform between localized fiducial markers and nominal electron-beam lithography (EBL) coordinates. To account for potential relative rotation, scaling, and translation between coordinate frames, we compute an optimal least-squares similarity transform between the encoder and image-based point clouds. In the absence of distortions and systematic errors, these coordinates are expected to align within calibration accuracy and encoder repeatability limits. However, we observe that even after optimal alignment (neglecting optical distortion corrections~\cite{copeland_traceable_2024}), the residual displacements ($r$) remain comparable to or larger than the characteristic nearest-neighbor spacing of emitters. Figure~\ref{fig:SI_encoder}(b) shows the aligned coordinates and residual vectors, where color indicates residual magnitude and unit vectors indicate direction. To quantify the impact on navigability, we consider the distribution of residual magnitudes normalized by the mean nearest-neighbor distance, $d_{NN}$, shown in Fig.~\ref{fig:SI_encoder}(c)). The average ratio is $\langle|r|/d_{NN}\rangle = 2.26$ with a standard deviation of 0.9. This standard deviation serves as a descriptive statistic of the physical spread of the alignment offsets across the sampled emitters, rather than a formal measurement uncertainty. This average ratio exceeds unity, the threshold past which positional errors become comparable to emitter spacing and assignment ambiguity is unacceptably high. Figure~\ref{fig:SI_encoder}(d) shows the residual component distributions and reveals no strong systematic bias. Even for inter-field navigation, shifts of several micrometers may be large enough to require a final field centering step that relies on our template-matching approach.

\section{Navigation Fidelity}
\begin{figure}[h!]
    \centerline{
    \includegraphics[width=1.0\textwidth]{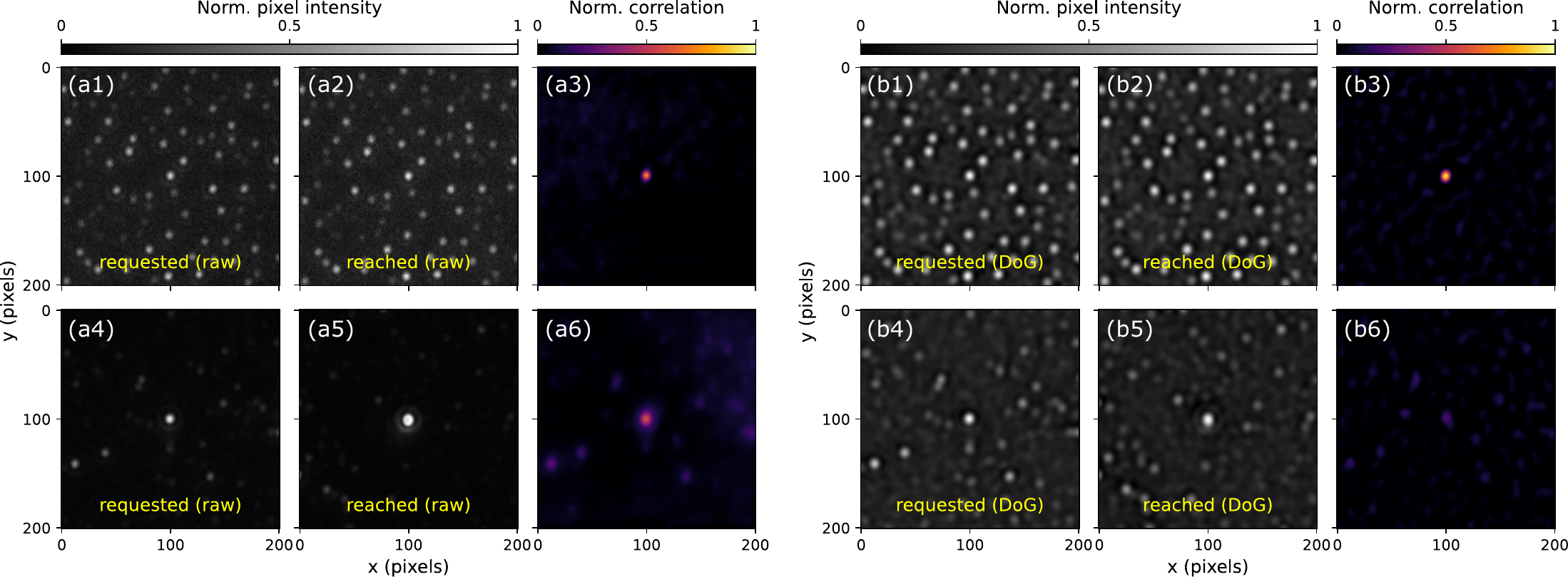}}
	\caption{
    Cross-correlation–based navigation verification. (a) Verification correlation using un-processed images. The top row (a1–a3) shows a successful (true-positive) navigation, while the bottom row (a4–a6) shows an incorrect (false-positive) navigation arising from intensity-dominated correlations. (a1, a4) are the original \textit{requested} ROIs; (a2, a5) are their respective \textit{reached} ROIs extracted around the confocal spot; (a3, a6) are the corresponding positive-only cross-correlations. (b) Same examples as in (a), but with both templates pre-processed using power-law intensity compression ($\gamma=0.3$) and DoG filtering ($\sigma_1=1.2\sigma_{\mathrm{PSF}}$ and $\sigma_2=1.6\sigma_1$). Here, the true-positive maximum correlation is improved from 0.78 to 0.97, indicating an improved degree of similarity between templates and a successful navigation. Similarly, the false-positive maximum correlation was reduced from 0.58 to 0.22, indicating a more accurate, reduced degree of similarity between templates. 
	} 
	\label{fig:SI_navigation}
\end{figure}

Since encoder-based navigation and verification can be unreliable at the smaller inter-PSF distance scales, we implement a final verification cross-correlation to quantify navigation confidence. That is, to determine if the ROI positioned at the confocal excitation/collection spot is the same one that was requested. Once navigation has converged, a final \textit{reached} ROI is extracted from the wide-field PL image, centered on the confocal spot. This final template is matched against the original \textit{requested} ROI to assess their similarity. We find that the raw cross-correlation can be unreliable, particularly for navigation snapshots acquired with short exposure time (i.e., lower signal-to-noise ratio (SNR)), or when using templates with low PSF density. This can potentially produce a false positive, where the correlation peak occurs at the incorrect position, and the wrong ROI is positioned at the confocal spot. This typically occurs when a single high-intensity PSF overlap can dominate the correlation (see Fig.~\ref{fig:SI_navigation} (a4-a6)).

To avoid this issue, gamma compression is applied to adjust image contrast by rescaling the pixel intensities following $I=I_0^{\gamma}$, which enhances dim features while suppressing high-intensity contributions. A Difference-of-Gaussians (DoG) filter is then applied to both correlation inputs, defined as the difference between two Gaussian-blurred images with standard deviations $\sigma_1$ and $\sigma_2$ ($\sigma_1<\sigma_2$). This operation acts as a band-pass filter, suppressing low- and high-frequency background components while enhancing structures with PSF length-scales. In combination, this pre-processing suppresses intensity-dominated correlations and emphasizes structural similarity, which greatly improves the reliability while preserving reasonable navigation speed. A comparison between raw and pre-processed ROIs is provided in Fig.~\ref{fig:SI_navigation}. The final correlation maximum is used as a navigation confidence score, which can be used to flag potential false positives during navigation to exclude from measurement or analysis. The distribution of these similarity scores is shown in Fig.~\ref{fig:SI_nav_stats}(c), highlighting the reliability of our image-based navigation.

\section{System performance and throughput}
\begin{figure}[h!]
    \centerline{
    \includegraphics[width=1.0\textwidth]{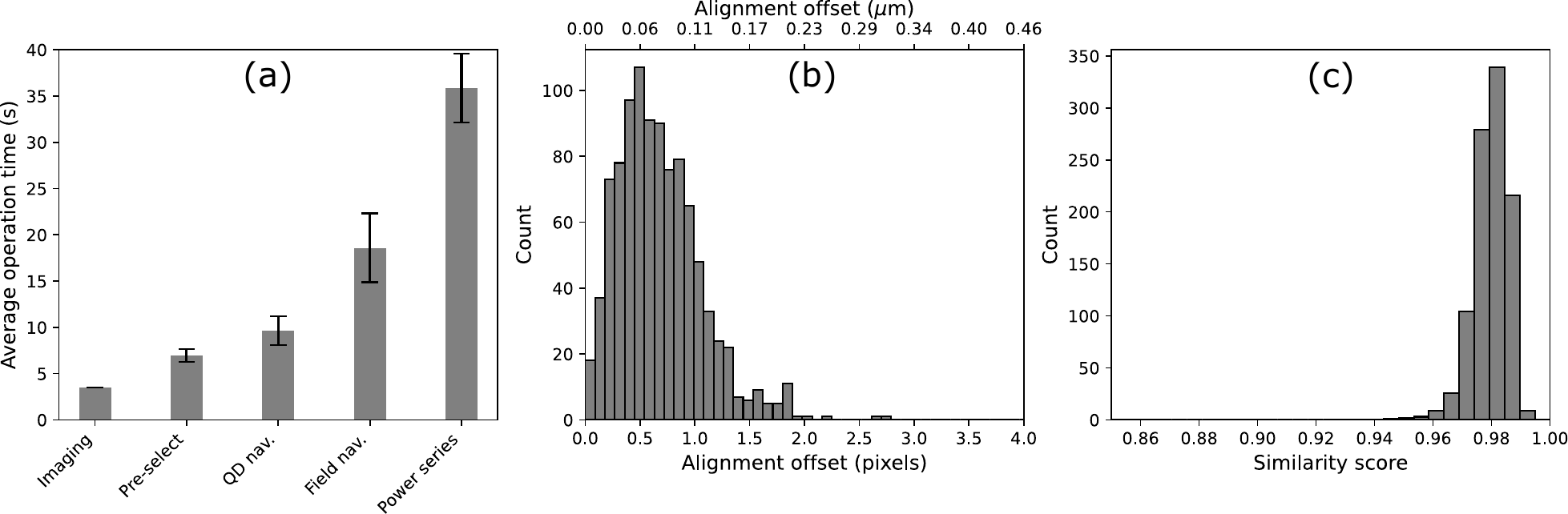}}
	\caption{Process timing and navigation performance for 18 separate fields of view and 988 total emitters. (a) Average time for each process from the automated measurement. Vertical lines represent the sample standard deviation to illustrate the variation in process timing. From left to right the processes are: \textit{imaging} -- acquiring wide-field PL and fiducial images; \textit{pre-select} -- running the full emitter pre-selection described in Fig.~\ref{fig:SI_selection}; \textit{QD nav.} -- navigating to each PSF; \textit{Field nav.} -- navigating between fields of view; \textit{Power series} -- acquisition of spectral power series (20 powers with \qty{1.5}{\second} integration time each). The variation in power series acquisition time is from compiling results from datasets using different power series acquisition parameters, and the variation in field navigation is likely from typewrite-style field translation (i.e., navigating from the end of one row to the start of another). (b) Final alignment offsets between fitted PSF center and confocal laser spot. The average PSF-confocal alignment accuracy is 0.69 pixels (\qty{79.1}{\nano\meter}) with a sample standard deviation of 0.5 pixels (\qty{57.1}{\nano\meter}), obtained using a scale factor from system calibration. This standard deviation characterizes the empirical spread of the system's spatial alignment performance. (c) Final navigation validation cross-correlation maxima showing a high average correlation maximum of 0.980 with a standard deviation of 0.006, provided as a descriptive metric of the cross-correlation spread, which indicates successful navigation to all requested emitters.
	} 
	\label{fig:SI_nav_stats}
\end{figure}

The total experimental runtime $t_{\mathrm{total}}$ can be estimated as
\begin{equation}\label{eq:throughput}
    t_{\mathrm{total}} = N_{\mathrm{fields}}(t_{\mathrm{nav}}^{\mathrm{field}}+\sum_{i=1}^{N_{\mathrm{bpf}}}(t_{\mathrm{image}, i}+t_{\mathrm{select}, i}+N_{\mathrm{psf}, i}(t_{\mathrm{nav}}^{\mathrm{psf}}+t_{\mathrm{spectra}}))),
\end{equation}
where $N_{\mathrm{fields}}$ is the total number of fields of view to visit, $t_{\mathrm{nav}}^{\mathrm{field}}$ is the average time to navigate between fields, $N_{\mathrm{bpf}}$ is the number of imaging band-pass filters used, $t_{\mathrm{image}}$ and $t_{\mathrm{select}}$ are the average times to acquire final wide-field images and pre-select emitters, respectively, $N_{\mathrm{psf}}$ is the number of PSFs to measure in each field (the summation over BPFs is included to account for PSF density variation between imaging filters), $t_{\mathrm{nav}}^{\mathrm{psf}}$ is the average time to navigate between PSFs, and $t_{\mathrm{spectra}}$ is the average spectral acquisition time, which will depend on number of spectral series, steps per series, and integration time per step. Ultimately, the characterization depth is configurable: the system can rapidly acquire a single spectrum per emitter to minimize total runtime, or perform extended spectral series for more detailed characterization at the cost of proportionally longer acquisition times. Measured process times are provided in Fig.~\ref{fig:SI_nav_stats}(a), and typical experimental use cases are modeled in Table~\ref{table:SI_throughput} to highlight the system's flexibility.

\begin{table}[h]
\centering
\begin{tabular}{|c|c|c|c|c|c|c|}
\hline
    & $N_{\mathrm{fields}}$ & $N_{\mathrm{bpf}}$ & $N_{\mathrm{psf}}$ & $t_{\mathrm{spectra}}$ (s) & total emitters & runtime (hh:mm:ss)\\
\hline
case 1 & 81 & 1 & 10 & 2 & 810 & 03:29:15\\
\hline
case 2 & 81 & 1 & 10 & 30 & 810 & 09:47:15\\
\hline
case 3 & 81 & 1 & 10 & 60 & 810 & 16:32:15\\
\hline
case 4 & 7 & 3 & 140 & 30 & 2940 & 32:47:35\\
\hline
\end{tabular}
\caption{
Estimated runtime for different measurement cases using the throughput model in Equation~\ref{eq:throughput} with fixed $t_{\mathrm{nav}}^{\mathrm{field}}=\qty{20}{\second}$, $t_{\mathrm{nav}}^{\mathrm{psf}}=\qty{10}{\second}$, $t_{\mathrm{image}}=\qty{5}{\second}$, and $t_{\mathrm{select}}=\qty{10}{\second}$, extracted from the performance summary shown in Figure~\ref{fig:SI_nav_stats}(a). Cases 1--3 correspond to a measurement of a single $9\times9$ grid of fields for different characterization depths. Case 1 represents a measurement solely for deterministic fabrication where only a quick spectrum is acquired for 10 emitters per field using a single BPF. Case 2 and 3 increase the measurement depth, for example, by including power- and polarization-series, respectively (each requires approximately \qty{30}{\second} additional measurement time). Case 4 is representative of the data shown in the main text, where power-series data was acquired for roughly 140 emitters per BPF across seven fields (140 is an average PSF count across BPFs for simplicity, which yields a similar runtime as treating each BPF separately).
}\label{table:SI_throughput}
\end{table}

It is important to highlight the flexible operation of our system and emphasize the tradeoffs between total runtime, spatial coverage, and information efficiency. For example, the pushbroom hyperspectral imaging (HSI) approach detailed by Buchinger \textit{et al.}~\cite{buchinger_deterministic_2025-1} estimates a measurement time of \qty{30}{\minute} per field of view (355 scan steps with \qty{3}{\second} integration time) to acquire single-power spectra across all pixels. Operating in a similar rapid-screening mode (single-power spectra, \qty{3}{\second} integration per emitter), our system can measure approximately 140 emitters in the same time frame. Depending on how many emitters are desired per field of view, those emitters may all be located in one field of view or spread over many fields of view, while in the same time period the HSI approach would have measured a single field of view. Compared to manual operation, we estimate our system to be roughly 10 to 50 times faster depending on operator experience and dedication.

Although longer screening times are expected for larger QD populations, especially for high-density QD populations where HSI systems have a throughput advantage, our selective measurement framework provides improved data efficiency for single-emitter datasets. The HSI approach generates roughly \qty{1.3}{\giga\byte} of data per field for a single-power hyperspectral image~\cite{buchinger_deterministic_2025-1}. This larger data overhead arises because HSI must construct a dense 3D datacube by scanning a magnified image of a field across a spectrometer slit (or by rastering a beam or stage while collecting spectra), thereby measuring potentially empty space between sparse emitters. In contrast, our vector-scanning approach bypasses this spatial oversampling by utilizing wide-field imaging to directly pre-localize and target only viable QDs. As a result, our system can acquire both baseline and full multi-dimensional spectral series using a fraction of the data. For a standard single-spectrum acquisition for 140 emitters in a single field of view, our approach consumes roughly \qty{130}{\mega\byte}---an order of magnitude reduction compared to the estimated \qty{1.3}{\giga\byte} required for a single slit-based HSI scan. However, this efficiency becomes far more pronounced when scaling to multidimensional datasets. For example, acquiring power- and polarization-dependent spectral series (e.g., 20 powers and 25 half-wave plate angles) for the same 140 emitters produces roughly \qty{140}{\mega\byte} per field of view. To acquire this equivalent multidimensional spectral and polarization information, we estimate an HSI setup would produce nearly \qty{60}{\giga\byte} of raw data per field. Thus, for multidimensional characterization, our approach achieves over a \qty{99}{\percent} reduction in data footprint (roughly $400\times$ less data). This massive compression ensures that mapping hundreds of fields on a single chip remains practically viable without requiring careful data cleaning.

\section{Bimodal Spectral Structure Clustering}\label{sec:SI_clustering}
\begin{figure}[h!]
    \centerline{\includegraphics[width=1.0\textwidth]{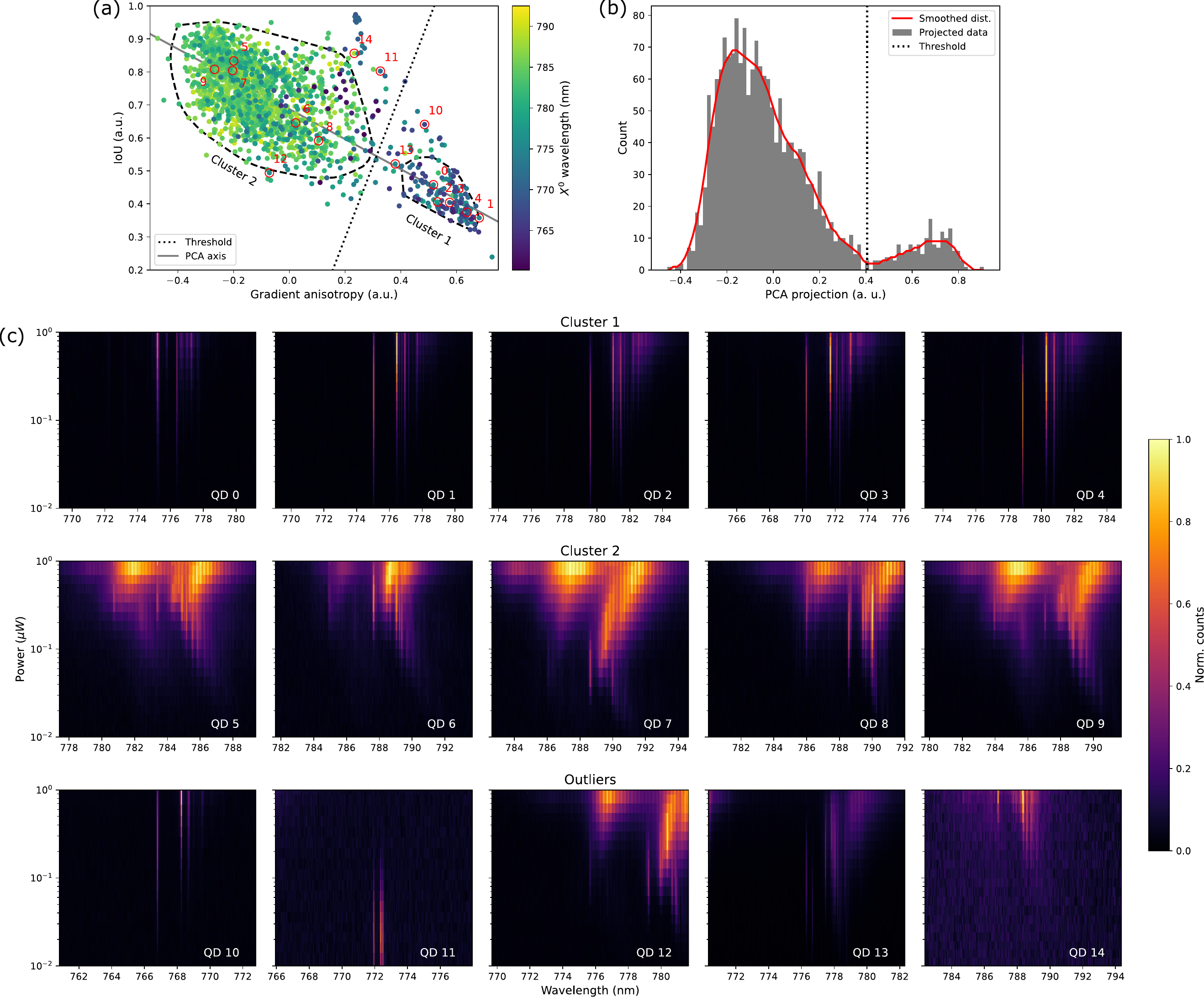}}
	\caption{Unsupervised machine-learning clustering analysis of QD spectral power series. (a) Distribution of intersection-over-union (IoU) and gradient score classification metrics colored by $\mathrm{X}^0$ wavelength. The dashed outlines represent the convex hull of each identified cluster (smoothed for visualization). Definitions and explanations for these metrics are described in Equations~\ref{eq:grad} to~\ref{eq:iou}. The HDBSCAN clustering algorithm, used with a minimum cluster size of 30, identifies two primary classes of spectra (labeled clusters 1 and 2; points outside the boundaries are considered outliers). Randomly selected spectra from each group are shown in (c), qualitatively illustrating the structural differences between them. A principal component analysis (PCA) is performed to identify the direction of maximal variance (PCA axis), which is used to define a decision threshold as described in (b). (b) Distribution of clustering metrics after projection onto the PCA axis. A decision boundary to separate high- and low-quality spectra is chosen as the minimum of the density valley between the two populations. (c) Randomly selected spectral power series from each cluster and outlier category in (a). The top row is from cluster 1, the middle row is from cluster 2, and the bottom row is from the outliers.
    }
	\label{fig:SI_clustering}
\end{figure}
A preliminary inspection of the collected power-dependent QD spectra suggested distinct spectral behaviors for QDs with emission above and below roughly \qty{780}{\nano\meter} (\qty{1.59}{\electronvolt}). QDs emitting below \qty{780}{\nano\meter}, captured with a \qty{770}{\nano\meter} bandpass filter, generally exhibited the characteristic signatures of desirable LDE GaAs dots~\cite{huber_single-particle-picture_2019}. Their spectra were dominated by a sharp neutral exciton peak on the high-energy side of a small cluster of narrow lines separated by approximately \qty{1}{\milli\electronvolt}, corresponding to charged and excited excitonic states. This profile remained stable over most of the selected pump power range, as seen in the top row of Fig.~\ref{fig:SI_clustering}(c). In contrast, QDs above \qty{780}{\nano\meter} wavelengths generally presented variable characteristics. Many of these dots featured only broad emission bands, as opposed to sharp lines, even at low excitation powers. Others exhibited sharp lines initially, but with less pronounced neutral exciton transitions which were quickly overtaken in intensity by lower-energy lines and bands as pump power increased. The low-energy lines were often less well resolved and tended to coalesce into a single broad emission band even at low power. Overall, red-shifting bands were observed with increasing pump powers, as seen in the middle row of Fig.~\ref{fig:SI_clustering}(c). 

To verify whether these characteristics were indeed shared among QD sub-populations above and below approximately \qty{780}{\nano\meter}, we used a hierarchical DBSCAN (HDBSCAN) algorithm~\cite{campello_density-based_2013, mcinnes_hdbscan_2017} to identify classes of power-dependent behavior. Raw spectral power series images are large ($N_{\mathrm{powers}}\times1024$ pixels). Searching for patterns across more than 20,000 individual data points per sample can overwhelm clustering algorithms with noise and obscures underlying trends, making dimensionality reduction necessary. However, standard reduction approaches (e.g., principal component analysis (PCA) or autoencoders) are sensitive to the absolute position of a peak on the x-axis, which can trivially group the data by its emission wavelength rather than its actual power-dependent behavior~\cite{ramsay_functional_2005}. To avoid this, we reduce the spectral images into two wavelength-agnostic metrics that purely describe the structural shape of the power series. By intentionally hiding absolute wavelength information from the algorithm, we ensure that the resulting clusters are driven entirely by structure, allowing us to evaluate how those behaviors correlate with wavelength independently afterward.


We first define a gradient-based metric, which captures anisotropy in 2D power-dependent spectra such as in Fig~\ref{fig:SI_clustering}(c), as:
\begin{equation}\label{eq:grad}
\mathrm{score} = \frac{G_{\lambda} - G_{P}}{G_{\lambda} + G_{P}},
\end{equation}
where $G_x$ is the root-mean-square (RMS) of the directional gradient magnitude in the photoluminescence (PL) intensity $Z(P,\lambda)$ along wavelength ($\lambda$) or excitation power ($P$):
\begin{equation}
G_x = \sqrt{\left\langle \left(\frac{\partial Z(P,\lambda)}{\partial x}\right)^2 \right\rangle}.
\end{equation}
Physically, $G_{\lambda} > G_{P}$ indicates stronger localization of features along the wavelength axis, while $G_{\lambda} < G_{P}$ reflects power-dominated behavior (e.g., power-dependent spectral broadening). 

To complement this, a secondary structure metric is constructed from the collapsed one-dimensional spectrum
\begin{equation}
\tilde{Z}(\lambda) = \frac{\sum_{P} Z(P,\lambda)}{\max_{\lambda} \sum_{P} Z(P,\lambda)}.
\end{equation}
An intersection-over-union (IoU) style metric is then computed over the envelopes of two threshold-response curves:
\begin{equation}
\mathrm{IoU} =
\frac{\int_0^1 \min\big(W(\theta), G(\theta)\big)\, d\theta}
{\int_0^1 \max\big(W(\theta), G(\theta)\big)\, d\theta},
\end{equation}
where $W(\theta)$ is a spectral support curve defined as the fraction of wavelength bins (out of a total $N$ bins) exceeding a threshold $\theta$,
\begin{equation}
W(\theta) = \frac{1}{N} \sum_{\lambda} \mathbf{1}\big(\tilde{Z}(\lambda) \ge \theta\big),
\end{equation}
and $G(\theta)$ is a spectral mass curve defined as the fraction of total collapsed intensity that survives the same threshold:
\begin{equation}\label{eq:iou}
G(\theta) =
\frac{\sum_{\lambda} \tilde{Z}(\lambda)\,\mathbf{1}\big(\tilde{Z}(\lambda) \ge \theta\big)}
{\sum_{\lambda} \tilde{Z}(\lambda)}.
\end{equation}
Here, $\mathbf{1}(x)$ is the indicator function, which equals 1 when $x$ is true and 0 when $x$ is false. The IoU quantifies the overlap between these two curves, capturing how spectral support and intensity mass co-evolve under increasing threshold. In particular, high IoU values correspond to spectra in which intensity is distributed uniformly across the supported wavelength range, whereas low IoU values indicate spectra with concentrated or structured emission. 

These two metrics, computed for all valid, collected power-dependent QD data, are plotted in Fig.~\ref{fig:SI_clustering}(a) and reveal a bimodal distribution in spectral power-series behavior in which two primary clusters (indicated by dashed boundaries) are initially identified with an HDBSCAN clustering algorithm. The outliers often have similar features to the spectra in either cluster, but are considered outliers primarily due to the minimum cluster size used in the HDBSCAN algorithm, and we rely on clustering purely as a coarse grouping method. 

We define a decision threshold from the local minimum of the data projected onto the first principal component axis (Fig.~\ref{fig:SI_clustering}(b)). As illustrated by representative spectral power series in the top row of Fig.~\ref{fig:SI_clustering}(c), emitters on the right side of the threshold are classified as ``good'' due to their favorable spectral characteristics, while those on the opposite side are classified as ``bad.'' Importantly, the ``good'' emitters predominantly correspond to emitters with $\mathrm{X}^0$ wavelengths below \qty{780}{\nano\meter}. Though the classification is obtained through wavelength-agnostic clustering, it is consistent with the behavior observed in the FWHM-wavelength correlation presented in the main text.


\section{PSF intensity statistics}
The full distribution of $\mathrm{X}^0$ linewidth and wavelength across the range of PSF pixel intensities, which were summarized in the main text, are provided here for completeness.

\begin{figure}[H]
    \centerline{\includegraphics[width=0.95\textwidth]{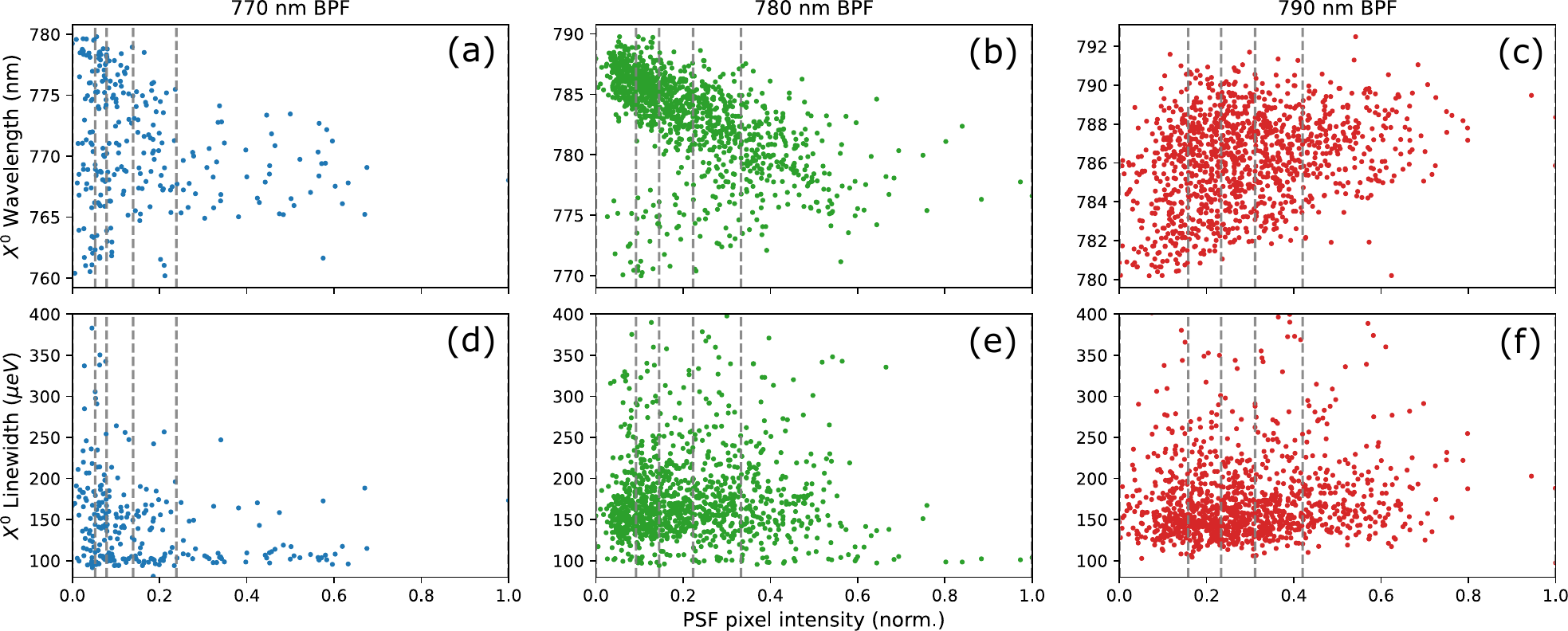}}
	\caption{Scatter plots relating spectral properties with PSF pixel intensity from emitters measured with different band-pass filters (BPFs). The intensities are taken simply as the pixel value at each PSF's centroid and are normalized to $[0,1]$ for visualization. The quintile bin edges are indicated by the dashed lines. (a-c) $\mathrm{X}^0$ emission wavelength versus PSF pixel intensity, showing that brighter PSFs generally have an emission wavelength closer to the center of the BPF pass-band. (d-f) $\mathrm{X}^0$ FWHM versus PSF pixel intensity, revealing minimal correlations apart from the \qty{770}{\nano\meter} dataset, where the brightest PSFs tend to have narrower linewidths.
    }
	\label{fig:SI_intensity_corr}
\end{figure}

\bibliography{references_Ned_v2, references_MD}